\documentclass[10pt,twocolumn,amsmath,amssymb,aps,prl,superscriptaddress,citeautoscript,longbibliography]{revtex4-2}

\usepackage{graphicx}
\usepackage{dcolumn}
\usepackage{bm}

\usepackage{xcolor}
\usepackage{soul}
\usepackage{graphicx}
\usepackage{dcolumn}
\usepackage{bm}
\usepackage{mathrsfs}
\usepackage{amsmath}
\usepackage{amssymb}
\def\lesssim{\ \raise.3ex\hbox{$<$}\kern-0.8em\lower.7ex\hbox{$\sim$}\ }
\def\gesim{\ \raise.3ex\hbox{$>$}\kern-0.8em\lower.7ex\hbox{$\sim$}\ }

\begin{document}
\title{Emergent Weyl Nodes and Berry Curvature in Bose Polarons \\ via $p$-Wave Feshbach Coupling}
\author{Hiroyuki Tajima}
\affiliation{Department of Physics, Graduate School of Science, The University of Tokyo, Tokyo 113-0033, Japan}
\affiliation{RIKEN Nishina Center, Wako 351-0198, Japan}
\affiliation{Quark Nuclear Science Institute, The University of Tokyo, Tokyo 113-0033, Japan}
\author{Eiji Nakano}
\affiliation{Department of Mathematics and Physics, Kochi University, Kochi 780-8520, Japan}
\author{Kei Iida}
\affiliation{Department of Liberal Arts, The Open University of Japan, Chiba 261-8586, Japan}
\affiliation{RIKEN Nishina Center, Wako 351-0198, Japan}

\date{\today}
\begin{abstract}
  We show that an impurity quasiparticle immersed in a Bose-Einstein condensate, known as a Bose polaron, exhibits topological properties characterized by
  a nonzero Berry curvature, which is induced by
Weyl nodes that emerge via interspecies $p$-wave Feshbach resonance.
Such nodes occur even in the absence of spin degrees of freedom and spin-orbit coupling.
For charged impurities, the corresponding $p$-wave polarons are shown to be accompanied by
chiral anomaly.
The above predictions can be tested in a cold atomic environment by observing the Hall transport of the atomic or ionic impurity cloud.
\end{abstract}
\maketitle

{\it Introduction}---
Topology plays a crucial role in various fields of modern physics,
ranging from
condensed-matter physics~\cite{RevModPhys.83.1057}, 
quantum chemistry~\cite{bradlyn2017topological},
photonics~\cite{RevModPhys.91.015006},
quantum computations~\cite{stern2013topological}
to high-energy physics~\cite{witten1988topological}.
For a deeper understanding of topological aspects in quantum many-body systems in such fields,
ultracold atoms provide an ideal testing ground thanks to their remarkable controllability.
Indeed, ultracold atoms act as quantum simulators of real systems.  For example, lattice systems can be optically prepared~\cite{schafer2020tools},
while a strong-coupling regime can be realized by using Feshbach resonances~\cite{RevModPhys.82.1225}. 
In the presence of bosonic and fermionic atoms,  moreover, one can set up various many-body states such as Bose-Einstein condensation~\cite{RevModPhys.71.463} and Fermi superfluidity~\cite{RevModPhys.80.1215}.
So far, topological properties in ultracold atoms have been examined by utilizing
optical lattices~\cite{RevModPhys.91.015005} and artificial gauge fields~\cite{goldman2014light}.

Once quantum simulation with ultracold atoms is set up, it is then essential to probe
many-body effects in such quantum systems.  
To this end, polarons, that is,
impurities surrounded by a cloud of medium excitations, attract much attention.
While the notion of polarons was originally introduced for electrons moving in an ionic crystal~\cite{landau1933electron,landau1948effective}, nowadays, experimental realizations in ultracold atoms provide us with not only a theoretical benchmark due to its simplified setup, but also
dramatic progress in our understanding of the relevant
thermodynamics~\cite{PhysRevLett.122.093401,yan2020bose,PhysRevX.10.041019,baroni2024mediated}
and nonequilibrium dynamics~\cite{skou2021non,PhysRevX.15.021070,vivanco2025strongly},
from which one can probe the properties of the medium.
In ultracold atomic mixtures, polarons are categorized as Fermi polarons and Bose polarons, where impurity atoms are embedded in Bose and Fermi media,
respectively, and theoretically studied for both cases~\cite{massignan2014polarons,scazza2022repulsive,tajima2021polaron,tajima2024intersections,PhysRevA.110.030101,grusdt2025impurities}.

Despite such progress, atomic polaron states exhibiting topological properties have yet to be realized or even theoretically studied, except for the cases where the medium itself has a non-trivial topology~\cite{PhysRevA.99.033613,PhysRevB.99.081105,vashisht2025chiral}. 
In particular, in the absence of optical lattices, a promising way to impose topological properties on gas systems
relies on artificial gauge fields, which, however, induce an unwelcome heating effect due to photon scattering~\cite{RevModPhys.83.1523}.

\begin{figure}[t]
    \centering
    \includegraphics[width=\linewidth]{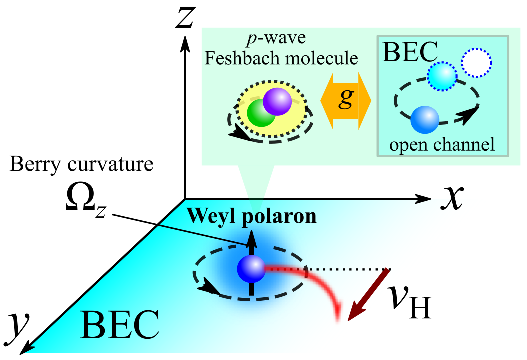}
    \caption{Schematic picture of a Weyl polaron in a Bose-Einstein condensate near the boson-impurity $p$-wave Feshbach resonance. The $p$-wave coupling $g$ induces the coherent dressing of the impurity state and the Berry curvature $\bm{\Omega}=(\Omega_x,\,\Omega_y,\,\Omega_z)$, leading to the anomalous velocity $v_{\rm H}$ just like the anomalous Hall effect.}
    \label{fig:1}
\end{figure}

In this Letter, we theoretically propose a straightforward yet unprecedented method for realizing a topological polaron state
without resorting to artificial gauge fields or topological media.
Specifically, we consider an impurity atom interacting with medium bosonic atoms via the $p$-wave Feshbach resonance,
that is, a $p$-wave Bose polaron~\cite{PhysRevC.111.025802}, as illustrated in Fig.~\ref{fig:1}. 
The
$p$-wave Feshbach resonance, which has its energy level split between the different values of
the $z$-component of the orbital angular momentum, i.e.,
$\ell_z=+1$ and $\ell_z=-1$, gives rise to the Weyl points and hence nonzero Berry curvature 
within the polaron state, even in the absence of a topological medium. 
Indeed, several $p$-wave Feshbach resonances are known to undergo splitting of the $\ell_z=-1$ and $\ell_{z}=+1$ levels
due to the electromagnetically anisotropic structure of the atoms involved,
as observed in $^6$Li-$^{133}$Cs mixtures~\cite{zhu2019spin} and in $^6$Li-$^{53}$Cr mixtures~\cite{PhysRevLett.129.093402}.
The orbital angular momentum of a Feshbach molecule acts to invoke the chiral structure of polaronic excitations, which was not explored in earlier studies of $p$-wave polarons~\cite{PhysRevC.111.025802,PhysRevLett.109.075302,PhysRevA.100.062712}.
It is also noteworthy that
the present scheme can be realized in an ion-atom mixture~\cite{weckesser2021observation,PhysRevX.15.011051}, which could be used as a reference system to explore chiral kinetic theories in the context of chiral magnetic effects~\cite{PhysRevLett.109.162001,PhysRevLett.109.181602,PhysRevD.87.085016,PhysRevLett.113.182302}.

{\it Theoretical model for $p$-wave Bose polarons}---
In what follows, we use $\hbar=1$ and assume a unit volume for convenience.
We consider a system in which impurity atoms are
immersed in an atomic Bose-Einstein condensate and interact with the medium atoms via
$\ell_z$-dependent $p$-wave Feshbach coupling with strength $g_{\ell_z}$ and splitting of the energy levels $\nu_{\ell_z}$
of the Feshbach molecules.
The corresponding two-channel Hamiltonian is given by
\begin{align}
\label{eq:1}
    H&=\sum_{\bm{k}}
    \left[\xi_{\bm{k},b}b_{\bm{k}}^\dag b_{\bm{k}}
    +
\xi_{\bm{k},c}c_{\bm{k}}^\dag c_{\bm{k}}
    +
    \sum_{\ell_z=0,\pm1}
    \xi_{\bm{k},A,\ell_z}{A}_{\bm{k},\ell_z}^\dag {A}_{\bm{k},\ell_z}
    \right]
    \cr
        &\,+\frac{1}{2}
    \sum_{\bm{k},\bm{k}',\bm{P}}
    U_{bb}
    b_{\bm{k}+\frac{\bm{P}}{2}}^\dag
    b_{-\bm{k}+\frac{\bm{P}}{2}}^\dag
    b_{-\bm{k}'+\frac{\bm{P}}{2}}
    b_{\bm{k}'+\frac{\bm{P}}{2}}\cr
    &\,+\sum_{\bm{k},\bm{P},\ell_z}
    \left[g_{\ell_z}{k}_{\ell_z}
    {A}_{\bm{P},\ell_z}^\dag b_{-\bm{k}+\frac{M_b}{M_A}\bm{P}}
    c_{\bm{k}+\frac{M_c}{M_A}\bm{P}}
    +{\rm h.c.}
    \right],
    \end{align}
where $\xi_{\bm{k},b}=k^2/2M_b-\mu_b$,
$\xi_{\bm{k},c}=k^2/2M_c-\mu_c$,
and $\xi_{\bm{k},A,\ell_z}={k^2}/{2M_A}+\nu_{\ell_z}-\mu_b-\mu_c$ are the kinetic energies of a Bose atom with mass $M_b$ and chemical potential $\mu_b$,
an impurity atom with mass $M_c$ and chemical potential $\mu_c$, and a closed-channel molecule with mass $M_A=M_b+M_c$, respectively, and
$b_{\bm{k}}^{(\dag)}$, $c_{\bm{k}}^{(\dag)}$, and $A_{\bm{k},\ell_z}^{(\dag)}$ are the corresponding annihilation (creation) operators; we do not specify the statistics
of the impurity atom and the closed-channel molecule.
In Eq.~\eqref{eq:1}, the second term is the repulsive boson-boson interaction with coupling constant $U_{bb}$, and the last term describes
the above-mentioned
$p$-wave Feshbach coupling
where $k_{\ell_z=\pm 1}=\mp{(k_x\pm ik_y)}/{\sqrt{2}}$ and $k_{\ell_z=0}=k_z$.
While $U_{bb}$ is necessary to keep the stability of the background condensate, the residual impurity-impurity and impurity-boson interactions are not important for our purpose and thus neglected for simplicity.

We
describe the Bose-Einstein condensate as a macroscopically occupied zero-momentum state via  
    $b_{\bm{k}}\rightarrow \sqrt{n_0}\delta_{\bm{k},\bm{0}}+\pi_{\bm{k}}(1-\delta_{\bm{k},\bm{0}})$,
where $n_0$ is the condensate density. 
Since $n_0\gg 1$, we can obtain the leading-order effective Hamiltonian of polarons as
\begin{align}
    H_{\rm LO}
    &=
    \sum_{\bm{p}}
    \Psi_{\bm{p}}^\dag
    \left(
    \begin{array}{cccc}
       \xi_{\bm{p},c}  & \Lambda_{\bm{p},+1}^* & \Lambda_{\bm{p},0}^* & \Lambda_{\bm{p},-1}^*  \\
       \Lambda_{\bm{p},+1}  & \xi_{\bm{p},A,+1} & 0 & 0 \\
       \Lambda_{\bm{p},0} & 0 &\xi_{\bm{p},A,0} & 0 \\
       \Lambda_{\bm{p},-1} & 0 & 0 & \xi_{\bm{p},A,-1}
    \end{array}
    \right)\Psi_{\bm{p}}
\end{align}
where $\Psi_{\bm{p}}=(c_{\bm{p}},\, A_{\bm{p},+1},\, A_{\bm{p},0},\, A_{\bm{p},-1})^{\rm T}$ is the vector that represents the polaronic states having momentum $\bm{p}$
  in open and closed channels,
and $\Lambda_{\bm{p},\ell_z}=\frac{M_b}{M_A}\sqrt{n_0}g_{\ell_z}p_{\ell_z}$ is the Rabi-like one-body coupling induced by the $p$-wave impurity-medium Feshbach coupling.
Hereafter, $g_{\ell_z}$ is taken to be
real
without loss of generality.

We focus on the case where the system is close to $\ell_z=+1$ resonance with $\xi_{\bm{p},c}\simeq \xi_{\bm{p},A,+1}$ at small $\bm{p}$ but
$|\xi_{\bm{p},A,0}-\xi_{\bm{p},c}|\gg |\Lambda_{\bm{p},0}|$ and $|\xi_{\bm{p},A,-1}-\xi_{\bm{p},c}|\gg |\Lambda_{\bm{p},-1}|$ (i.e., the magnetic field is tuned to near the $\ell_z=+1$ resonance).
The off-resonant $\ell_z=0$ and $-1$ channels give only higher-order corrections
suppressed by the inverse square of the energy difference~\cite{RevModPhys.82.1959} and can be neglected to leading order. 
In such a case, the low-energy effective Hamiltonian reduces to
\begin{align}
\label{eq:3}
    H_{\rm LO}^{\rm eff}
    &=\sum_{\bm{p}}
    \psi_{\bm{p}}^\dag
    \left(
    \begin{array}{cc}
       \xi_{\bm{p},c}  & \Lambda_{\bm{p},+1}^*   \\
       \Lambda_{\bm{p},+1}  & \xi_{\bm{p},A,+1}  \\
    \end{array}
    \right)
    \psi_{\bm{p}}\cr
    &\equiv\sum_{\bm{p}}\psi_{\bm{p}}^\dag 
    [\bm{\sigma}\cdot\bm{d}(\bm{p})+d_0(\bm{p})]
    \psi_{\bm{p}},
\end{align}
where $\psi_{\bm{p}}=(c_{\bm{p}},\, A_{\bm{p},+1})^{\rm T}$ and $\bm{d}(\bm{p})=\left(
-\zeta p_x
,\, 
-\zeta p_y
,\, \frac{\xi_{\bm{p},c}-\xi_{\bm{p},A,+1}}{2}\right)$ with 
$\zeta=\frac{M_b\sqrt{n_0}g_{+1}}{\sqrt{2}M_A}$
and
$d_0(\bm{p})=(\xi_{\bm{p},c}+\xi_{\bm{p},A,+1})/2$ are the effective spinor and the $d$-vector, respectively.
Here, $\bm{\sigma}$ is the Pauli matrix that acts on $\psi_{\bm{p}}$.

Even in the absence of spin degrees of freedom,
the Weyl nodes in the momentum space are inherent in Eq.~\eqref{eq:3}; 
$\bm{d}(\bm{p}=(0,\,0,\,\pm p_{\rm W}))=\bm{0}$ is satisfied at the band-crossing momentum $p_{\rm W}$ given by
\begin{align}
    p_{\rm W}=\sqrt{\frac{\nu_{+1}-\mu_b}{\frac{1}{2M_c}-\frac{1}{2M_A}}}\theta(\nu_{+1}-\mu_b).
\end{align}
Since $\partial_{p_z}d_z(\bm{p})|_{\bm{p}=(0,\,0,\,p_{\rm W})}=\left(\frac{1}{2M_c}-\frac{1}{2M_A}\right)p_{\rm W}\neq 0$ for nonzero $p_{\rm W}$,
    we can see 
    that $\bm{p}_{{\rm W},\chi}=(0,\,0,\,\chi p_{\rm W})$ is indeed a Weyl point with chirality $\chi=\pm 1$ as
    $\bm{\sigma}\cdot\bm{d}(\bm{p})\propto 
    \chi \bm{\sigma}\cdot(\bm{p}- \bm{p}_{{\rm W},\chi})$ around $\bm{p}=\bm{p}_{{\rm W},\chi}$ after rescaling and unitary transformation.

Accordingly, we obtain the Berry connection
\begin{align}
    \bm{\mathcal{A}}_{\bm{p}}^{\kappa}=i\langle u_{\kappa}(\bm{p})|\nabla_{\bm{p}}|u_\kappa(\bm{p})\rangle
\end{align}
and the Berry curvature
\begin{align}
    \bm{\Omega}^{\kappa}(\bm{p})=\nabla_{\bm{p}}\times\bm{\mathcal{A}}_{\bm{p}}^{\kappa}
\end{align}
for the normalized eigenstate $|u_{\kappa=\pm}(\bm{p})\rangle$ with the eigenenergy $ E_{\kappa=\pm}(\bm{p})=\kappa|\bm{d}(\bm{p})| +d_0(\bm{p})$.
It is useful to express $\bm{d}(\bm{p})$
in terms of the polar coordinate
as $\bm{d}(\bm{p})=|\bm{d}(\bm{p})|(\sin\theta_{\bm{p}}\cos\phi_{\bm{p}},\sin\theta_{\bm{p}}\sin\phi_{\bm{p}},\cos\theta_{\bm{p}})$,
leading to 
\begin{align}
\label{eq:7}
    \bm{\mathcal{A}}_{\bm{p}}^{\kappa}=-\kappa\frac{1-\cos\theta_{\bm{p}}}{2}\nabla_{\bm{p}}\phi_{\bm{p}}.
\end{align}
Note that $\mathcal{A}_{\bm{p}}^\kappa$ depends on the gauge choice. 
In order to avoid the Dirac string, it is necessary to introduce two patches and perform a gauge transformation in each region~\cite{wu1976dirac}.
However, the Berry curvature does not depend on the gauge transformation and is given by
\begin{align}
        \bm{\Omega}^{\kappa}(\bm{p})=-\kappa\frac{\sin\theta_{\bm{p}}}{2}\nabla_{\bm{p}}\theta_{\bm{p}}\times\nabla_{\bm{p}}\phi_{\bm{p}}.
\end{align}
In particular, the $z$-component of $\bm{\Omega}^\kappa(\bm{p})$ reads
\begin{align}
    \Omega_{z}^{\kappa}(\bm{p})=\kappa\frac{\zeta^2}{2|\bm{d}|^3}
    (d_z-p_x\partial_{p_x}d_z-p_y\partial_{p_y}d_z).
   \label{eq:9}
\end{align}

\begin{figure*}[t]
    \centering
    \includegraphics[width=\linewidth]{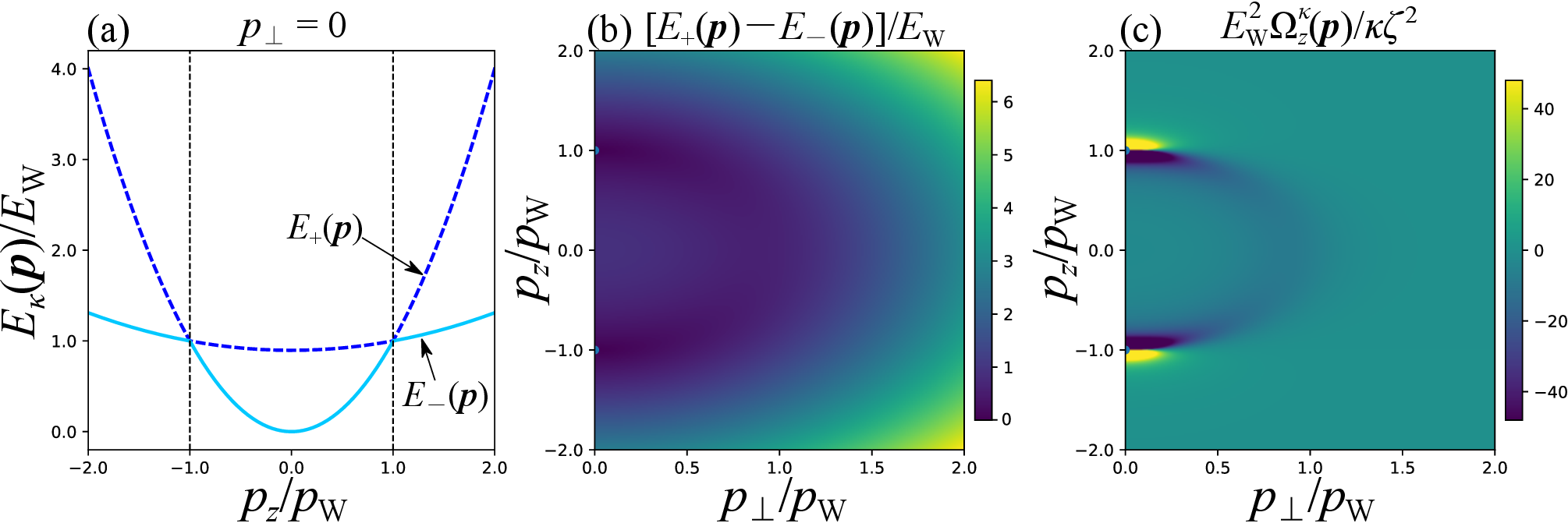}
    \caption{(a) Polaron dispersion $E_{\kappa=\pm}(\bm{p})$ at $\bm{p}=(0,\,0,\,p_z)$,
    where the solid (dashed) curves represent the ground (excited)-state branch. 
    Energy gap (b), $E_+(\bm{p})-E_{-}(\bm{p})$, and the
    Berry curvature (c), Eq.\ (\ref{eq:9}), are displayed in the plane of $p_\perp$ and $p_z$, where $E_{\rm W}=p_{\rm W}^2/2M_c$ is used for the normalization. In these figures, we employ $2M_c\zeta^2/E_{\rm W}=0.1$, $\mu_{c}=0$, and $M_c/M_b=6/52$.}
    \label{fig:2}
\end{figure*}

Figure~\ref{fig:2}(a) shows the polaron dispersion $E_{\kappa}(\bm{p}=(0,\,0,\,p_z))$
with $M_c/M_b=6/52$, a value relevant to a $^6$Li-$^{52}$Cr mixture~\cite{PhysRevLett.129.093402}, and $2M_c\zeta^2/E_{\rm W}=0.1$.
At $p_z=\pm p_{\rm W}$, level crossing occurs between two dispersions with different $\chi$, which have
different curvatures in a manner that is reminiscent of the bare dispersions given by $p^2/2M_c$ and $p^2/2M_{A}$.
Note that the quasiflat dispersion is associated with the closed-channel molecular dispersion.
In this sense, the present atomic mixture with large mass imbalance
could be used as a quantum simulator
of topological flat bands in multi-band systems~\cite{aoki2025flat}.

To see the region with nonzero $p_x$ and $p_y$, as shown in Fig.~\ref{fig:2}(b),
we plot the energy gap $E_{+}(\bm{p})-E_{-}(\bm{p})$ between the polaron excited and ground states
in the plane of $p_\perp=\sqrt{p_x^2+p_y^2}$ and $p_z$.
The energy gap is found to close at the Weyl points $\bm{p}_{{\rm W},\chi}=(0,\,0,\,\chi p_{\rm W})$ and increases as $\bm{p}$ goes away from these points. 
Figure~\ref{fig:2}(c) shows the
Berry curvature $\Omega_z^{\kappa}(\bm{p})$ in the $z$-direction as given by Eq.\ (\ref{eq:9}).
Since the Weyl points located at $\bm{p}=(0,\,0,\,\pm p_{\rm W})$ work as the Dirac monopoles in the momentum space,
$\Omega_{z}^{\kappa}(\bm{p})$ diverges at these points with opposite chiralities $\chi=\pm 1$.
We emphasize that
such a non-trivial Berry curvature
is attributable to the $p$-wave resonance and the background condensate,
a feature crucially different from well-known condensed matter systems involving spin-orbit coupling~\cite{RevModPhys.82.1959} and anisotropic pairing order~\cite{sato2017topological}.

While we confine ourselves to the case where the system is close to the $p$-wave Feshbach resonance with $\ell_z=+1$, we can likewise consider the case of $\ell_z=-1$, where we find
$\bm{\Omega}^{\kappa}(\bm{p},\ell_z=-1)=-\bm{\Omega}^{\kappa}(\bm{p},\ell_z=+1)$.
This reflects the fact that the present Berry curvature is related to the level splitting of the $p$-wave Feshbach resonances.
Also, while we demonstrate the case of a $^6$Li-$^{52}$Cr Bose-Fermi mixture in Fig.~\ref{fig:2}, 
similarity of the scattering properties with the well-documented case of a $^{6}$Li-$^{53}$Cr Fermi-Fermi mixture has also been discussed~\cite{PhysRevLett.129.093402}, which suggests that no correction is required except the reduced mass.
Thus, our proposal holds not only for
bosonic impurities but also for fermionic impurities.
To see the clear signature of the emergent Berry curvature as a first step, fermionic impurities would be more useful
because the Pauli-blocking effect would suppress the effect of mediated interactions.

{\it Semiclassical equation of motion and anomalous velocity}---
To see the effect of the Berry curvature on a mobile topological Bose polaron,
we derive the corresponding semiclassical equation of motion~\cite{RevModPhys.82.1959}.
An important advantage of the present system lies in the feasibility of measuring its microscopic dynamics.
The anomalous velocity due to the Berry curvature can be measured by observing the center-of-mass dynamics of the polaron cloud~\cite{yan2024collective}.
The effective action is given by
\begin{align}
    S=\int dt \langle \Psi|i\partial_t-H_{\rm LO}^{\rm eff}|\Psi\rangle\equiv \int dt \, L,
\end{align}
where $|\Psi\rangle$ is the state vector for a wave packet of the polaron with center position $\bm{r}_{\rm c}$ and momentum $\bm{p}_{\rm c}$. 
Within the semiclassical approximation~\cite{RevModPhys.82.1959},
we can obtain the effective Lagrangian as
\begin{align}
    L\simeq \dot{\bm{r}}_{\rm c}\cdot{\bm{p}}_{\rm c}+\dot{\bm{p}}_{\rm c}\cdot\bm{\mathcal{A}}_{\bm{p}_{\rm c}}^{\kappa}-E_{\kappa}(\bm{p}_{\rm c}). 
\end{align}
The Euler-Lagrange equation for $\bm{p}_{\rm c}$ leads to
the semiclassical equation of motion
\begin{align}
\label{eq:12}
    \dot{\bm{r}}=\nabla_{\bm{{p}}}E_{\kappa}(\bm{p})-\dot{\bm{p}}\times\bm{\Omega}^{\kappa}(\bm{p}),
\end{align}
where we suppress the index ${\rm c}$ for convenience.
The second term on the right side of Eq.~\eqref{eq:12} is the {\it anomalous velocity}, which is responsible for the Hall transport of polarons.

Meanwhile, under the external potential $U(\bm{r})$,
polarons feel the driving force $\bm{F}=-\bm{\nabla} U(\bm{r})$ as the second equation of motion is given by $\dot{\bm{p}}=\bm{F}$.
Eventually, we obtain
\begin{align}
    \dot{\bm{r}}
    =\nabla_{\bm{{p}}}E_{\kappa}(\bm{p})-\bm{F}\times\bm{\Omega}^{\kappa}(\bm{p}).
\end{align}
This expression indicates that the impurity cloud moves in the direction
perpendicular to $\bm{F}$ and $\bm{\Omega}^{\kappa}(\bm{p})$, leading to the anomalous velocity $\bm{v}_{\rm H}$.
For example, in the case of
$\bm{F}=(F_x,0,0)$ and sufficiently small momenta,
we can obtain the anomalous Hall velocity in the $y$-direction as
\begin{align}
    v_{{\rm H},y}\simeq F_x\Omega_{z}^{\kappa}(\bm{0})\equiv\kappa\frac{\zeta^2}{(\nu_{+1}-\mu_b)^2}F_x,
\end{align}
indicating that the anomalous velocity is enhanced near the $p$-wave resonance.
Note that the present semiclassical expression for $v_{{\rm H},y}$ diverges
at the resonance, in the vicinity of which the low-momentum semiclassical approximation breaks down.
A more qunatitative analysis would require allowance for a fully quantum nature near the resonance.

By using the semiclassical equation of motion,  moreover, we can obtain the bulk mass current of
topological Bose polarons with $\bm{F}=(F_x,0,0)$ as
\begin{align}
    \bm{j}&=m\sum_{\bm{p}}
    \sum_{\kappa=\pm}
    f(E_{\kappa}(\bm{p}))\left[\nabla_{\bm{p}}E_{\kappa}(\bm{p})-\bm{F}\times\bm{\Omega}^{\kappa}(\bm{p})\right]\cr
    &=-m\sum_{\kappa=\pm}\int \frac{d^d\bm{p}}{(2\pi)^d}
    [\bm{F}\times\bm{\Omega}^{\kappa}(\bm{p})]f(E_{\kappa}(\bm{p})),
\end{align}
where we use the fact that the momentum integration of $f(E_{\kappa}(\bm{p}))\nabla_{\bm{p}}E_{\kappa}({\bm{p}})\equiv \nabla_{\bm{p}}g(E_{\kappa}({\bm{p}}))$ gives zero. 
Accordingly, $\Omega_{y}^{\kappa}\propto p_yp_z$ leads to $j_z=0$.
On the other hand, Eq.~\eqref{eq:9} gives nonvanishing $j_y$ as
\begin{align}
    j_y=F_x\sum_{\kappa=\pm}\int \frac{d^d\bm{p}}{(2\pi)^d}
    {\Omega}_{z}^{\kappa}(\bm{p})f(E_{\kappa}(\bm{p})),
\end{align}
where $f(x)=1/(e^{\beta x}\pm 1)$ is the Fermi/Bose distribution function of impurities.
The mass Hall conductivity can thus be written as
\begin{align}
    \sigma_{yx}=
    \frac{j_y}{F_x}=
    \sum_{\kappa=\pm}\int \frac{d^d\bm{p}}{(2\pi)^d}
    {\Omega}_{z}^{\kappa}(\bm{p})f(E_{\kappa}(\bm{p})),
\end{align}
which can also be extracted from the center-of-mass motion of the impurity cloud~\cite{sommer2011universal,sommer2011spin}.
This expression suggests that
our system could provide an opportunity to investigate both
fermionic and bosonic anomalous Hall transport by changing the impurity
statistics.
For fermionic impurities at $T=0$ and $\mu_c=E_{\rm W}$,
we obtain $\sigma_{yx}=p_{\rm W}/2\pi^2$ as in the cases of
Weyl semimetals~\cite{RevModPhys.90.015001} and
dense QCD matter~\cite{tatsumi2018anomalous}.

Finally, we briefly mention the case
in which atomic impurities are replaced by
charged impurities, e.g., ionic  polarons~\cite{astrakharchik2021ionic}. 
In such a case, an impurity with
charge $q$ feels both the electric field $\bm{E}$ and the magnetic field $\bm{B}$ through
$\dot{\bm{p}}=q(\bm{E}+\dot{\bm{r}}\times\bm{B})$.
Substituting it into the semiclassical equation of motion,
we obtain 
\begin{align}    \dot{\bm{p}}=\frac{q\bm{E}+q\nabla_{\bm{p}}E_{\kappa}(\bm{p})\times\bm{B}-q^2(\bm{E}\cdot\bm{B})\bm{\Omega}^{\kappa}(\bm{p})}{1-q\bm{B}\cdot\bm{\Omega}^{\kappa}(\bm{p})},
\end{align}
leading to the chiral anomaly,
$\chi \frac{q^2}{4\pi^2}\bm{E}\cdot\bm{B}$, in the continuity equation for the chirality-resolved quasiparticle density~\cite{PhysRevLett.109.181602,PhysRevD.87.085016}.
In this regard, charged $p$-wave Bose polarons have the potential to simulate the chiral magnetic effect~\cite{kharzeev2014chiral}. 

{\it Summary}---
We theoretically elucidate a topological nature of atomic
polarons immersed in an atomic Bose-Einstein condensate near the $p$-wave Feshbach resonance.
This polaron state exhibits the Berry curvature induced by the $p$-wave coupling with the background condensate, which plays a role in
coherent mixing between impurity atoms and closed-channel molecules.
In contrast to the conventional ideas that utilize polarons
as a probe of a non-trivial medium,
polarons themselves are predicted to acquire the emergent Berry curvature in the presence of the mixing-induced pseudospin structure.
Our prediction can be tested by measuring the Hall transport of polaron clouds in ultracold atom experiments.

For future developments, many interesting problems remain.
For example, the mediated interaction between fermionic impurities is expected to induce $p$-wave Fermi superfluidity~\cite{PhysRevB.65.134519,PhysRevA.77.043629,PhysRevLett.121.253402,PhysRevA.108.023304}, which, in the present scheme, is accompanied by intrinsic Berry curvatures and thus could be used for the quantum simulation of topological flat-band superconductors~\cite{aoki2025flat}
without any lattice geometries. 
It is also interesting to see quantum geometry effects in the present system~\cite{yu2025quantum}.

{\it Acknowledgment.---}
The authors thank J. Takahashi, K. Ochi, and Nuclear theory group at Univ. Tokyo for
useful discussions.
This work is supported by JSPS KAKENHI for (Grants No. JP22K13981, No. JP23K22429, No.
JP23K25864, No. JP24K06925, and No. JP25K01001) from
MEXT, Japan.

\bibliographystyle{apsrev4-1}
\bibliography{reference.bib}

\end{document}